\documentclass[useAMS,usenatbib]{mn2e}
\usepackage{graphicx}

\usepackage{todonotes}
\def\lesssim{\mathrel{\hbox{\rlap{\hbox{\lower3pt\hbox{$\sim$}}}\hbox{\raise2pt\hbox{$<$}}}}}
\def\gtrsim{\mathrel{\hbox{\rlap{\hbox{\lower3pt\hbox{$\sim$}}}\hbox{\raise2pt\hbox{$>$}}}}}
\newcommand\ion[2]{#1$\;${\scshape{#2}}}

\title[Spectroscopic Boundaries of Type I Supernovae]{On the Spectroscopic Boundaries Between Normal and Peculiar Type~I Supernovae}
\author[J. T. Parrent]{J. T. Parrent$^{1, 2, 3}$\thanks{E-mail:
jparrent@cfa.harvard.edu (JTP)}\\
$^{1}$6127 Wilder Lab, Department of Physics \& Astronomy, Dartmouth College, Hanover, NH 03755, USA\\
$^{2}$Las Cumbres Observatory Global Telescope Network, Goleta, CA 93117, USA\\
$^{3}$Harvard-Smithsonian Center for Astrophysics, 60 Garden St., Cambridge, MA 02138, USA}
\begin{document}


\pagerange{\pageref{firstpage}--\pageref{lastpage}} \pubyear{2014}

\maketitle

\label{firstpage}

\begin{abstract}
The spectrum of a supernova is a summation of numerous overlapping atomic line signatures. Consequently, empirical measurements are limited in application when compound features are assumed to be due to one or two spectral lines. Here I address matters of spectroscopic boundaries between normal and peculiar type I supernovae using multi-component empirical metrics. I discuss some obstacles faced when using supernova spectra to pair model with data and I demonstrate how spectrum synthesis can benefit from fairly complete observational coverage in wavelength and time.
\end{abstract}


\section{Introduction}

The ability to associate spectral features with individual atomic line transitions allows for the direct measure of time-dependent, full-width half-maxima of various line profiles. It is this fundamental, model-independent association with a given set of lines that enables astronomers to iteratively test and constrain detailed abundance models thereby inferring previous stages of an observed phenomenon or spectral sequence \citep{Vreux96}.

For example, such procedures of direct analysis have been used to study the solar spectrum. Astronomers had documented the Sun's spectral features down to sub-Angstrom scales by the late 1960s, often to the second decimal place. Roughly a quarter of these lines spanning optical wavelengths were found to overlap, creating thousands of unresolved sets of two or more lines. Still, indirect measurements of each constituent profile could be obtained by leveraging constraint through the remaining unblended lines of the same atomic species. Had the same fraction of lines been unblended, there would still be 3,322 optical lines to leverage parameters for 89 atomic ionization states, with the remaining 1,612 lines associated with molecules in our own atmosphere and that of the Sun \citep{Moore66,Kurucz11}.

By contrast, the 10$^{4}$~km~s$^{-1}$ expansion velocities of expelled material in a supernova lead to severe blending of all lines; the spectrum of a supernova could be described as a 3-dimensional object (c.f.~\citealt{Blondin11a,Blondin12,Pauldrach14}). This alone ensures that isolated equivalent-widths of spectral signatures are not well-determined.


In terms of a taxonomy or spectral sequence, supernovae can be divided into two central categories, types~I~and~II, depending on whether or not the spectrum contains conspicuous signatures of hydrogen. For type~II supernovae, signatures of hydrogen Balmer lines are as apparent as they are for A-type stars \citep{Minkowski41}. 

In the case of thermonuclear type Ia supernovae (SN~Ia), conspicuous detections of hydrogen are thus far a mix of strong narrow emission juxtaposed with a weak P Cygni component of absorption (c.f.~\citealt{Aldering06,Prieto07,Dilday12,Silverman13IaCSM,Fox14,Maeda14}). For some core-collapse SN~Ib and SN~Ic events, conflicting estimates of projected Doppler velocities between signatures of \ion{Si}{ii} and \ion{Fe}{ii} suggest the outermost layers of ejecta may instead be contaminated with unburned hydrogen, or other trace signatures of progenitor material \citep{Anumpama05,Elmhamdi06,Parrent07,Ketchum08,James10,Hachinger12HHe,Dessart12,Milisavljevic14}. 


Much progress has been made toward understanding progenitors of hydrogen-poor SN~I \citep{McCully14b,Chrazov14,Fesen14}. However, explosive conditions and binary configurations remain unknown given that state of the art models have not yet reached a robust fluency capable of accounting for the full range of observed properties for ``nearest neighbor'' SN~I subtypes \citep{Branch07a,Jeffery07}. For SN~Ia, these include (1) SN~1994D-like Core Normal, (2) SN~1984A-like Broad Line, (3) luminous Shallow Silicon SN~1991T-likes, and (4) Super-Chandrasekhar Candidate SN~2003fg-likes (hereafter, SCC), in addition to (5) peculiar Cool, faint SN~1991bg and (6) SN~2002cx-like SN~Ia \citep{Howell01a,Howell06,Branch06,Branch09,Doull11,Narayan11,Scalzo12,Foley13,Stritzinger14,Stritzinger1412Z}.

As the spectrum of each SN~I subtype evolves over timescales of days, traceable differences for a given set of spectra are often too subtle or too ill-defined by low to moderate signal-to-noise ratios and incomplete follow-up coverage in both wavelength and time. When spectroscopic diversity among normal, peculiar, and ``twin'' SN~I events is primarily in terms of those self-similar, heterogeneous differences that are discernible above adjacent line-blending, an ability to pinpoint and measure continually red-shifting spectral features becomes non-trivial. 

In spite of severe line blending, there are a multitude of spectral signatures one encounters when inspecting SN~I spectra. For normal SN~Ia the list thus far includes but is not limited to lines shown in Table~1. Depending on which set of compound features host the strongest lines, any given ion can be said to provide a permeable boundary through P~Cygni summation at optical wavelengths if the number of inferable signatures is either greater than or equal to two (c.f.~\citealt{Jeffery90}). 

Consequently, when one or two atomic signatures are associated with compound spectral features (e.g., near 6100~\AA\ and 8200~\AA), the use of intensity minima as robust indicators for so-called emergence, cutoff, and detachment velocities can lead to either unresolved estimates and/or over-interpretations of underlying physical differences. Hence, details of spectroscopic sequences for SN Ia, SN Ib, and SN Ic have been difficult to establish \citep{Blondin12,Modjaz14}.


Here I discuss one solution toward mapping spectroscopic patterns (or boundaries) by treating direct measurement as a ``constrained nonlinear optimization problem'' \citep{ThomasSYNAPPS}, whereby minimization between synthetic and observed spectra amounts to an internally consistent application of a given empirical line-identification metric \citep{Branch07a,Parrent10}. However, it remains to be seen what level of accuracy is needed for comparative SN studies \citep{Kerzendorf14,Sasdelli14PCA}.

In \S2 I address several topics associated with the challenges and successes of direct analysis of SN~Ia spectra, while several considerations can be applied to the spectra of SN~Ib and SN~Ic as well. In \S3 I describe the functionality of the fast and highly parameterized line identification code, \texttt{SYN++} \citep{ThomasSYNAPPS}. In \S4 I outline applications of a simplified line identification metric and discuss where automated and data-driven spectrum synthesis tools, such as \texttt{SYNAPPS}, can be used to improve comparative studies of SN.\footnote{Visit https://c3.lbl.gov/es/ for more details.} In \S5 I review empirical quantifications of SN~Ia diversity and related topics, and in \S6 I summarize and highlight steps integral for extracting observables of SN~I spectra.

\begin{table}
 \centering
 \begin{minipage}{140mm}
  \caption{Common Lines in Optical Spectra of SN~Ia}
  \begin{tabular}{ll}
  \hline
Ion & Rest Wavelength (\AA) \\ 
\hline
\ion{C}{ii} & 6580, 7234 \\
\ion{O}{i} &7774 \\ 
\ion{Mg}{ii} & 4481, 7896 \\ 
\ion{Ca}{ii} & 3969, 3934, 8498, 8542, 8662 \\
\ion{Si}{ii} & 3838, 4130, 5051, 5972, 6355 \\
\ion{Si}{iii} & 4560, 5743 \\
\ion{S}{ii} & 4163, 5032, 5208, 5468, 5654, 6305 \\
\ion{Fe}{ii} & 4025, 4549, 4924, 5018, 5169 \\
\ion{Fe}{iii} & 4420, 5156, 6000 \\
\hline
\end{tabular}
\end{minipage}
\end{table}

\section{Analysis of Supernova Spectra}

Spectral features near 5750~\AA\ and 6100~\AA\ for SN~Ia are often assumed to be associated primarily with the evolution of \ion{Si}{ii} line signatures. However, for most SN~Ia caught prior to maximum light, these features are shaped by either non-standard distributions of intermediate-mass elements or distinctly separate regions of high velocity \ion{Si}{ii} above an ostensible photospheric region of line formation \citep{Mazzali05a,Maund10a,vanRossum12,Marion13}. 

So-called ``detached \ion{Si}{ii}'' may exhibit a prolonged detection in some events near maximum light \citep{Branch05,Parrent12,Childress13HVF,Maguire14}, while photospheric \ion{Fe}{ii} and \ion{Fe}{iii} have been shown to influence the global minimum and wings of the 6100~\AA\ feature throughout the subsequent evolution \citep{Bongard08}. In short, the canonical \ion{Si}{ii} features most often used as spectroscopic probes of physical diversities can be shaped by a blend of asynchronous and interdependent line signatures from other species.

In fact, apart from spectroscopic differences accrued from higher expansion velocities and/or shallower radial opacity profiles, peculiar variability seen at optical wavelengths is rooted in the overall shapes and relative strengths of select compound features blueward of 6500~\AA. For example, between two or more SN~Ia, it is unusual for pseudo-equivalent widths of 5750~\AA\ features to approach that for 6100~\AA\ features; it is peculiar when the ratios of these pseudo-equivalent widths approach either zero or unity \citep{Nugent95a,Hatano00,Bongard06,Bongard08}. It is also unusual when spectral features are relatively narrow and weak \citep{WLi03,Howell06,Hicken07,Stritzinger14}. 

Occasionally the appearance of a spectrum is dominated by higher ionization states of iron \citep{Flipper91T,Jeffery92,Mazzali95,Hatano02}. For other events, low ionization titanium lines either weakly imprint sharp signatures \citep{Branch04c,Silverman13SN2013bh} or blanket larger wavelengths regions blueward of 5500~\AA\ \citep{Flipper92,Turatto98,Modjaz01,Garnavich04,Doull11}.

Given a lack of conspicuous boundaries for any given line or ion, an abundance of empirical false-positives is readily available within an underspecified parameter space of ``detections.'' Hence, studies of SN samples benefit when the set of priors, or in most cases the number of candidate lines, is either effectively maximized or close to the number of lines that are indeed present \citep{Baron96,Friesen14}. 

For SN~Ia, these are important points to consider when interpreting their spectra given that freshly synthesized ejecta remain contaminated by potentially unburned and distinguishable traces of C, O, Si, Ca, and Fe-rich progenitor material near and above the primary volume of line formation (a so-called ``photosphere,'' see \citealt{HatanoAtlas,Branch05}). It follows that an observed spectral signature associated with a particular ion at photospheric velocities (PV) can be influenced by identical ion and/or other atomic species at similar or relatively higher velocities (HV;~\citealt{Marion13}) and vice versa. A consequence is that detections and non-detections of various ions and spectral lines do not follow classical definitions; e.g., a non-detection of the strongest line from an ion at 5500~\AA\ does not necessarily imply a lack of significant spectral contamination from an ion at 5500~\AA\ and other wavelengths.

\subsection{Spectral Line Identification}

A detection of a line from an ion, including associated spectral sequences from one event to another, provides quantifiable traits of normal and peculiar phenomena \citep{Kromer13SN2010lp}. However, for those weak signatures of unburned progenitor material and/or high velocity components, a detection of a line from an ion necessitates a representative and complete set of hypotheses for the whereabouts of all other spectral lines. 

Singly ionized silicon and sulfur are considered detected in SN~Ia optical spectra when a number of spectral lines are conspicuous above noise and adjacent blending. However, detections of species with a single strong line (e.g., \ion{Na}{i}, \ion{C}{iii}, and \ion{Si}{iv}) are difficult to prove without a more detailed handling of line formation. 

For example, \citet{Dessart14} has argued against \ion{Na}{i}~D in the post-maximum spectra of SN~Ia. Based on NLTE calculations of delayed detonation models, and using a modified version of \texttt{CMFGEN} \citep{Hillier98} they find the 5900~\AA\ peak during post-maximum epochs is significantly shaped by [\ion{Co}{iii}] emission. Because freshly synthesized cobalt is found in abundance during these epochs, and despite densities that are a few orders of magnitude above critical densities, \citet{Dessart14} argue forbidden lines play an important role in shaping the 5900~\AA\ feature. 

Thus, the identity of the 5900 \AA\ emission feature as shaped by either \ion{Na}{i}~D or [\ion{Co}{iii}] remains ambiguous in spite of the fact that \ion{Na}{i}~D lines are either prevalent or identified in the spectra of various other SN~I events \citep{Dessart10,Dessart12}. Such ambiguity in the identity of a common element in SN~I spectra gives an indication of the difficulty of performing line identifications in general. 

Detections also serve to constrain models that do not similarly produce the same set of blended signatures over time. As an example, \citet{Sasdelli14} claim non-detections for all signatures of carbon in the early spectra of SN~1991T without providing an explanation for a tentative detection of \ion{C}{iii} $\lambda$4969; i.e., abundance determinations of carbon-rich regions would benefit from knowing the subset of species giving rise to weak 4500~\AA\ features in the spectra of SN~1991T/1999aa-like events \citep{Hatano02,Parrent11}.

\subsection{Line Identification Through Decomposition}

Usually, when an ion produces conspicuous signatures at a particular wavelength the parameters of associated lines can be deduced through spectral decomposition. Most inferences of an ion can be made relatively straightforward with a spectrum synthesis code when various sums of atomic signatures are compared or minimized to observations \citep{Thomas04,Silverman11,Foley13profiles,Hsiao13,Stritzinger1412Z,Chakradhari14}. 

Of course, such inferences should be compositionally consistent and match characteristics at other wavelengths. If this can be satisfied without an {\it ad hoc} prescription of available parameters, an ion has been inferred for the given metric and can be said to be provisionally detected. For example, often the sum of Mg, Si, S, and Ca indicate a lack of \ion{Fe}{ii} and other iron-peak ionization signatures like \ion{Ti}{ii}, \ion{Co}{ii}, and \ion{Ni}{ii} when compared to observations. Subsequently, optical detections of \ion{Mg}{ii} are difficult to constrain given that the strongest signatures are bounded by a list of ions including \ion{O}{i}, \ion{Si}{iii}, \ion{Fe}{ii}, and \ion{Fe}{iii}.

For SN~Ia, where spectral decompositions are more well-determined compared to those for SN~Ib and SN~Ic (c.f.~\citealt{Branch99,Modjaz14}), an empirical metric for one or two compound features that is based on one or two signatures from one or two ions allows for false-positives and certain errors in estimates of projected Doppler velocities (see \S5.4). One reason for this is that typical minimum resolutions of projected Doppler velocities are approximately 500~km~s$^{-1}$ on account of bulk line blending. 

Consider that an intrinsic difference of 500, 1000, and 2000~km~s$^{-1}$ would be reflected by a line-shift of $\sim$~10, 20, and 40~\AA, respectively for blended PV \ion{Si}{ii}~$\lambda$6355. That is, significant differences of 2000 km~s$^{-1}$ give rise to blueshifts that are of similar order to redshifts seen for 6100~\AA\ features during the first month of free expansion. When confronted by events like SN~2012fr \citep{Childress13,Maund13}, the length scales for consistency with compound features near 6100~\AA\ are relatively large. 

\subsection{Element Non-detection}

Lines of \ion{S}{i} are not reported in SN~Ia optical spectra. However, if the depressions near 6400~\AA\ and 7500~\AA\ happen to be well represented by \ion{S}{i} through spectrum synthesis of detailed explosion models, then it would follow that \ion{S}{i} could be inferred in some optical spectra. Confirming spectroscopic probes such as these  require joint optical$+$infrared data and analysis. Even so, the conclusions would remain statements of likelihoods for a given empirical prescription rather than firm detections.

Likewise, because \ion{S}{iii} has few strong optical signatures, this ion has not been detected in the spectra of Core Normal SN~Ia. However \ion{S}{iii} may contribute a non-negligible amount at blueward wavelengths in ``hot,'' Shallow Silicon SN~1991T/1999aa-like spectra \citep{Hatano02,Parrent11}. A corollary for the inference of \ion{S}{iii} in SN~1991T-like events is that \ion{S}{iii} may still provide some contribution to the same region in more normal SN~Ia spectra, which are thought to be slightly ``cooler'' in terms of the strongest ionization species observed \citep{Nugent95a}.

Another example of a non-detection is \ion{C}{ii} $\lambda$6580, which is not conspicuously detected in the spectra of many if not all Broad Lined SN~Ia, e.g., SN~2002bo \citep{Branch87,Lentz01b,Benetti04,Stehle05} and SN~2010jn \citep{Hachinger13}. In contrast for SN~2011fe-like events, the 6300~\AA\ and 6900~\AA\ features associated with \ion{C}{ii}~$\lambda$6580 and $\lambda$7234, respectively, suggest carbon exists down at projected Doppler velocities of $\sim$~8000~km~s$^{-1}$ as it finally fades from the spectrum as a conspicuously detectable line \citep{Pereira13}. 

While models can profit from such cutoff velocities, a non-detection for SN~2011fe-like events does not necessarily imply a lack of appreciable carbon below any last trace of 6300~\AA\ features \citep{Baron03}. Similarly, and at least for Core Normal and Shallow Silicon events, a non-detection of \ion{C}{ii}~$\lambda$6580 would not imply a symmetric distribution of carbon in the case of some merger scenarios (c.f.~\citealt{Pakmor12,Moll14}).

\subsection{Weak Features in SN~Ia spectra}

\begin{figure}
\centering
\includegraphics*[scale=0.55]{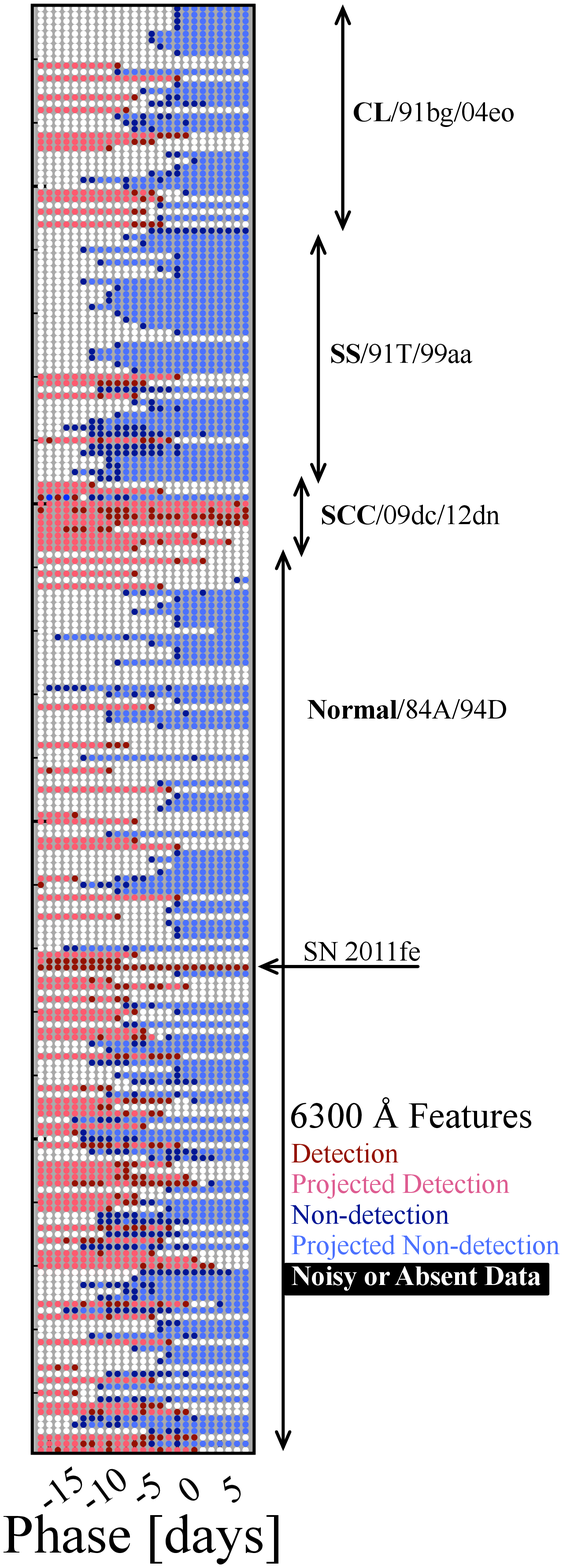}
\caption{Survey of \ion{C}{ii}~6300~\AA\ absorption features in terms of detections and non-detections for 228 SN Ia. Data are from \citet{Matheson08}, \citet{Parrent11}, \citet{Folatelli12}, \citet{Blondin12}, \citet{Silverman12spectra,Silverman12b}, \citet{Pereira13}, \citet{Folatelli13}, and \citet{Maguire14}. See legend for information on the color scheme; white dots represent no data taken or uncertainties due to low to moderate signal-to-noise ratios.}
\label{Fig:dotplot}
\end{figure}


The presence of weak spectral features can be deduced from inductive comparisons between adjacent epochs; i.e., to investigate weak, blended, and transient spectral features, e.g. those associated with \ion{C}{ii}~$\lambda\lambda$6580, 7234 absorption signatures, data needs are relatively high. This can be readily seen in Figure~\ref{Fig:dotplot} where I have plotted detections and non-detections of 6300~\AA\ absorption features for a selection of SN Ia spectra from current samples.  Here, an ideal centerpiece for a comparative survey is the normal SN~2011fe since its spectroscopic evolution was observed in exquisite detail \citep{Pereira13}. 

In particular, the early evolution of SN~2011fe's spectra revealed a redward moving signature from $\sim$~6300~\AA\ to 6400~\AA\ ($\sim$~4~\AA\ per day); the 6300~\AA\ feature in SN~2011fe advanced from one side of the underlying \ion{Si}{ii} $\lambda$6355 emission to the other just prior to maximum light (Fig.~9 of \citealt{Pereira13}). As SN~2011fe reached peak luminosity, the signal of the feature fell below noise and adjacent line-blending. 

Altogether this implies that conspicuous 6300~\AA\ signatures in the spectra of SN~Ia will be frequently detected blueward of the neighboring ``\ion{Si}{ii} emission'' \citep{Thomas07,Thomas11,Cartier14}. For normal SN~Ia similar to SN~2011fe, detections of weak 6300~\AA\ signatures near maximum light may also be found redward of the nearest emission feature; the blueward and redward extent of 6300~\AA\ absorption features are available for empirical measurement. One caveat is how to then interpret and measure the wavelength interval of weak detections \citep{Branch77,Jeffery90}. 

Because complete spectroscopic coverage is unobtainable at times, one can capitalize on two facts to enhance the informational return of the data. When a spectrum exhibits a 6300~\AA\ feature on day~$-$5, the feature is also present on day~$-$12; ``carbon-positive'' SN~Ia harbor 6300~\AA\ features during any unobserved epochs prior to the first detection. Similarly, a non-detection of a 6300~\AA\ feature on day~$-$5 implies a non-detection on day~$-$4 and thereafter. Again, the distinction of ``carbon-negative SN~Ia'' would not necessarily hold true despite a non-detection of a 6300~\AA\ feature \citep{Baron03,Blondin12,Hsiao13}. 

Thus, for select spectroscopic patterns, projections of detections and non-detections in phase space can alleviate the consequences of insufficient coverage and low signal-to-noise ratios. Assuming a thoroughly vetted and volume-limited sample of SN Ia, a summation over relative epochs might reveal clear patterns when paired with other observations, e.g. luminosities, colors, and/or properties of the host galaxies \citep{Pan15}. 

\section{Spectral Modeling: \texttt{SYNOW}/\texttt{SYN++}/\texttt{SYNAPPS}}

As a fast line identification tool, \texttt{SYN++} assumes spherical symmetry, homologous expansion, continuous emission from a sharp photosphere, and line formation treated in the Sobolev approximation. \texttt{SYN++}, formerly \texttt{SYNOW} \citep{Branch83,Branch85,Fisher00}, essentially views photospheric phase supernova spectra as a summation of resonance line P~Cygni profiles atop an underlying pseudo-continuum level. While spectrum formation is not similarly trivial for supernova atmospheres that are inherently time-dependent, non-local thermodynamic equilibrium plasmas, the one-dimensional \texttt{SYN++} model has served as a useful and complimentary tool alongside more detailed assessments of various spectroscopic quandaries.

Assuming a compositionally consistent set of constrainable ions, one can quickly evaluate observations; i.e. by hypothesizing the prescription responsible for each compound feature, a synthetic spectrum can be minimized to the data for the available parameter space.\footnote{For \texttt{SYN++}, this is most often \texttt{log~$\tau$}, \texttt{v$_{phot}$}, \texttt{v$_{min}$}, \texttt{aux}, and multiplied by the number of ions included.} In turn, this enables a tracing of a monotonically decreasing v$_{line}$ over time ($\pm$~$\delta$$v$; resolutions set by local line-blending). 


 Two of the top-most \texttt{SYN++} parameters are the relative location, \texttt{v$_{min}$}, and the relative strengths of the lines which can be set by \texttt{log~$\tau$}. Optical depths of the remaining lines assume LTE level populations, where the radial opacity profile declines outward as an exponential with an {\it e}-folding parameter near unity. 
 
 To facilitate the convergence of a fit, the continuum level is set by an assumed blackbody (\texttt{T$_{BB}$} in \texttt{SYNOW}; \texttt{t$_{phot}$} in \texttt{SYN++}). Spectral energy distributions of SN~I are not well-represented by blackbodies \citep{Bongard06,Bongard08} and the same can be said for SN~II \citep{Hershkowitz86approx,Hershkowitz86numerical,Hershkowitz87}. However, the assumption for \texttt{SYN++} is one of practicality and is not meant to be a reliable indicator of a real temperature structure. 
 
 In \texttt{SYN++}, quadratic warping constants \texttt{a$_{0}$}, \texttt{a$_{1}$}, and \texttt{a$_{2}$} supplement \texttt{t$_{phot}$} to form a more flexible continuum reference level; i.e., \texttt{a$_{0}$}, \texttt{a$_{1}$}, \texttt{a$_{2}$}, and \texttt{t$_{phot}$} form the backbone to data-driven minimizations and provide a more precise framework for direct analysis of supernova spectra. (See \citealt{ThomasSYNAPPS} and Fig.~2 of \citealt{Parrent12} for an example.) Thus, for well-observed events, one can conduct various experiments to directly assess: (i)~detectable and inferable lines of noteworthy ions, including C, O, Mg, Si, S, Ca, Fe, and Co; (ii) how to best utilize and constrain empirical metrics (\S4); (iii) the overlapping span of burned and remaining progenitor material in terms of projected Doppler velocities; (iv) the aforementioned emergence, cutoff, and detachment velocities; (v) the time-dependent prescription and related uncertainties for several compound features \citep{Scalzo14,Milisavljevic14}; and (vi) line-velocity gradients and plateaus (\S5.1).

\section{Application of Empirical Metrics}


In general, representative best fits with \texttt{SYN++} are initialized by exploring single and multi-ion comparisons while perturbing the available parameters. Because parameterized expansion velocities, $v_{line}$, are traced by blended atomic signatures over time, the convergence of \texttt{v$_{min,Y}$}(t) depends on the list of actively contributing atomic species used as input. However, since \texttt{SYN++} does not solve for ionization balance, which is a necessary step for abundance determinations and detailed analysis \citep{Baron96}, empirical metrics for \texttt{SYN++} are devised by perturbing compositional constructs \citep{HatanoAtlas}. 

Synthetic fits for a given set of ions can then be processed forward and backward in phase space relative to maximum light. This notion is one of an onion shell paradigm for the radiation transport in supernova atmospheres, which is innately limited. However, the \texttt{SYNOW} model can be used to constrain the composition structure of ejected matter despite being confined to a metric space \citep{Parrent10}.

A standard metric stems from \citet{Hatano99} and \citet{Branch05}, which is based on the time-dependent inclusion of \ion{C}{ii}, \ion{O}{ii}, \ion{Mg}{ii}, (PV~$+$~HV)~\ion{Si}{ii}, \ion{Si}{iii}, \ion{S}{ii}, (PV~$+$~HV)~\ion{Ca}{ii}, \ion{Cr}{ii}, (PV~$+$~HV)~\ion{Fe}{ii}, \ion{Fe}{iii}, and \ion{Co}{ii}. Based on the earliest spectrum of SN~2012dn, for example, it is reasonable to suspect there will be little need for the inclusion of HV \ion{Si}{ii} and HV~\ion{Ca}{ii} \citep{Chakradhari14}. However below I will discuss how this does not necessarily imply these HV structures are physically absent, nor that they should be left out of a more complete metric for higher levels of comparative analysis with automated \texttt{SYNAPPS}-like tools.

To ensure minimal under and over-shooting in $v_{min}$ for each spectrum, I re-examine the results for a range of saturated optical depths for each ion (c.f.~\citealt{Branch77}). Seeing that spectrum synthesis with \texttt{SYN++} is a minimization procedure, if $v_{min}$ for any ion is found to be off by more than 1000~km~s$^{-1}$, e.g., as is most often the case for \ion{Fe}{ii}, \ion{Fe}{iii}, and HV signatures, the full time-series minimization improves by perturbing $v_{min}$ by $\pm$~1000~km~s$^{-1}$ and repeating the above fitting methodology. In the event that the fitting procedure produces either monotonically increasing $v_{min, Y}$ over time, or oscillatory solutions with amplitudes exceeding 1000~km~s$^{-1}$, I interpret this as arising from either insufficient minimization, or the time-dependent presence of adjacent and unaccounted for lines, $v_{min, Z}(t)$, rather than real blueward shifts during a period of photospheric recession \citep{Branch73,Kirshner73}.\footnote{Ultimately, this may better explain empirically determined ($\Delta\lambda$/$\lambda_{0}$) indications of \ion{He}{i} layers below bulk regions of \ion{Fe}{ii} in the ejecta of some SN~IIb \citep{Ergon14,Folatelli14}. Here the use of $\Delta\lambda$/$\lambda_{0}$ assumes a correction for relativistic beaming has been made.}

\begin{figure}
\centering
\includegraphics*[scale=0.33]{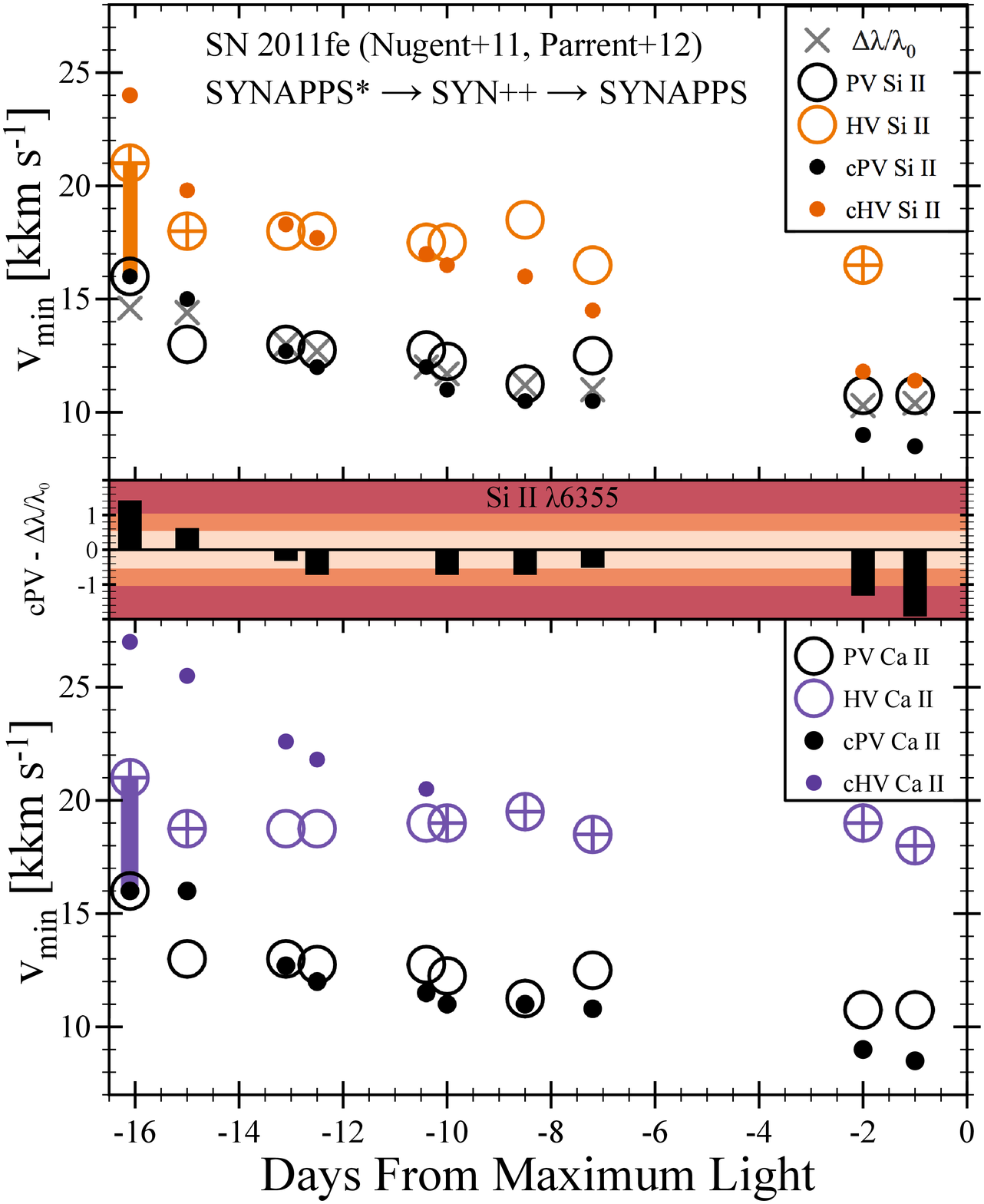}
\caption{Comparison between initial (*) and final \texttt{SYNAPPS} analysis for SN~2011fe \citep{Nugent11,Parrent12}. Shown are the initial and final values of ``converged'' \texttt{v$_{min}$}, prefixed here with the letter `c.' Units are in kkm s$^{-1}$~$\equiv$~km~s$^{-1}$~1000$^{-1}$. Crossed-circles indicate when the boundaries of the target compound feature are ill-defined and therefore uncertain for deduction of HV~\ion{Si}{ii} and HV~\ion{Ca}{ii} parameters; crossed-circles do not necessarily represent bonafide detections. Residuals betweens cPV and $\Delta$$\lambda$/$\lambda_{0}$ (without correcting for relativistic beaming) are shown in the middle panel for \ion{Si}{ii}~$\lambda$6355. See text in \S2 and \S3.}
\label{Fig:converge}
\end{figure}

Hence, I define ``best fits'' as satisfying the condition that a fit starting with iron-peak elements converge to a similar consensus in $v_{min}$-space for a fit ending with the inclusion of iron-peak elements (i.e., $v_{min}$ to within 1000~km~s$^{-1}$ for each ion considered). With approximately three to five seconds per output spectrum on average, i.e., assuming minimal computing resources, the above vetting process yields on the order of 1~$-$~2~x~10$^{3}$ synthetic spectrum comparisons before $v_{min,Y}(t)$ begins to converge toward a global minimum in the available parameter space given up by highly blended spectra. 

To demonstrate these considerations more clearly, I compare the initial and final \texttt{SYNAPPS} results for observations of SN~2011fe in Figure~\ref{Fig:converge}. The initial values of \texttt{v$_{min}$} for (PV~$+$~HV) \ion{Si}{ii} and \ion{Ca}{ii} on day $-$16 were taken from the previous \texttt{SYNAPPS} analysis of \citet{Nugent11}. For the remaining epochs, \texttt{v$_{min}$} for \ion{Si}{ii} and \ion{Ca}{ii} (PV and HV) were extrapolated through the above minimization procedure with \texttt{SYN++}. The full series of spectra were then processed with \texttt{SYNAPPS} until reaching a convergence in \texttt{v$_{min, Y}$}. Points represented by crossed-circles indicate when either of the compound features near 6100~\AA\ or 8200~\AA\ display a fully blended pair of PV~$+$~HV signatures. 

Note the discrepancy between initial and final \texttt{v$_{min}$}, spanning between $\sim$~500~$-$~4000~km~s$^{-1}$, depends on the phase of spectroscopic evolution and therefore the quality of a given dataset \citep{Patat96}. Remarkably, detachment velocities of HV~\ion{Si}{ii} near-maximum light and thereafter via direct analysis with \texttt{SYNAPPS} (small orange circles in Figure~\ref{Fig:converge}) are consistent with those found independently via spectropolarimetry \citep{Smith11,Parrent12}; i.e., direct analysis with \texttt{SYNAPPS} works and can be improved upon \citep{Friesen12}. However, the primary point of Figure~\ref{Fig:converge} is that the discrepancy in \texttt{v$_{min, Y}$}(t) between initial and final minimizations is a by-product of reaching local versus global minima in the metric space of \texttt{SYNAPPS}, which itself is a direct consequence of uncertainties incurred from line-blending alone. 

Thus, for incomplete datasets, estimates or ``direct measurements'' of projected Doppler velocities through either low or high-order empirical metrics are not well-determined upon initial inspection. Subsequently, when considering events that either share similar observed characteristics or border intrinsic divisions among distinct progenitor scenarios, comparisons between nearest-neighbor SN~Ia subtypes can become problematic. The same can be said for spectroscopic boundaries provided by the spectra of two homogenous SN~Ia (\S5.1).    


\section{Discussion}


Since the late 1980s, non-standard events have challenged our general understanding of candidate progenitor systems and explosion scenarios that give rise to the observed diversity of self-similar events \citep{Branch87}. Historically peculiar prototypes are often represented by SN~1984A, 1991T, 1991bg, and 2002cx, while so-called ``hybrid subtypes'' of SN~Ia are depicted by prototypes like SN~1999aa \citep{Garavini04}, 2001ay \citep{Krisciunas11,Baron12}, 2004dt \citep{Leonard05,Wang06}, 2004eo \citep{Pastorello07a}, 2006bt \citep{Foley10}, and PTF10ops \citep{Maguire11}.



Conceivably, those events representing extreme instances of SN~Ia, e.g. SN~1991bg and 2002cx, may in fact result from mechanisms or progenitor channels that are distinct from those most representative of the true norm of SN~Ia (\citealt{Hillebrandt13} and references therein). However a true norm could also originate from multiple binary configurations and explosive conditions \citep{Benetti05,Blondin12,Dessart14models}. Alternatively, select subclasses may either represent bimodal characteristics of one or two dominant channels \citep{Maedanature,Maund10a,Roepke12,Liu14}, or a predominant family where continuous differences such as progenitor metallicities (among other parameters) yield the observed diversity \citep{Nugent95a,Lentz00,Hoflich02,Scalzo14a,Maeda14}. A dominant one (or two) progenitor system~$+$~explosion mechanism, may then manifest properties about the norm that reach out into the same observational parameter space inhabited by the most peculiar subgroups \citep{Baron14}.

A more recent example of a SN~Ia with nominally conflicting descriptions is SN~2012dn \citep{2012dnCBET}. As was also the case for SN~2006gz \citep{Hicken07}, SN~2012dn is spectroscopically similar to super-Chandrasekhar mass candidates like SN~2007if and 2009dc \citep{Scalzo10,Silverman11,Kamiya12}. However both SN~2006gz and 2012dn are not as luminous as other members of the spectroscopic subclass \citep{Maeda09,Brown14,Chakradhari14}. In fact, given SN~2012dn's moderate luminosity and spectroscopic similarity to more standard events, it is important to distinguish those progenitor scenarios that lead to a similar set of observables without producing SN~2012dn-like events. 

In a forthcoming work, I will analyze the photospheric spectra of SN~2012dn with \texttt{SYN++} and compare the results alongside those obtained by \citet{Parrent12} for SN~2011fe (see \citealt{Chakradhari14} for a complementary analysis). However, to discuss SN~2012dn within the greater context of mapping spectroscopic boundaries, it is substantive to review the extent to which a spectroscopic sequence has been empirically established.

\subsection{Spectral Sequences}

Attempts to quantify observed patterns of SN~Ia at optical wavelengths have been made in the past. \citet{Branch06,Branch09}, \citet{Silverman12maxlight}, and \citet{Folatelli13} used pseudo-equivalent width measurements of canonical 5750~\AA\ and 6100~\AA\ spectral features during maximum-light epochs to quantify the peculiarities of SN~1984A-like, SN~1991T-like, and SN~1991bg-like events that set them apart from normal SN~1994D-like events. Essentially this particular use of pseudo-equivalent widths facilitates a consistent quantification of extreme and less conspicuous instances of peculiar variability from an uncertain and qualitatively defined norm.

SN~Ia diversity has also been categorized by the rate in change of the same 6100~\AA\ absorption minima during photospheric phases \citep{Benetti05,Blondin12,Dessart14models}, which has given rise to so-called ``high and low-velocity gradient'' SN~Ia (e.g., SN~1984A and SN~1994D, respectively). Here the assumption is one of a single line (\ion{Si}{ii}~$\lambda$6355) that manifests its signature as the observed and well-rounded absorption minimum for all SN~Ia subtypes during photospheric phases. However, this metric breaks down for SN~2012fr and Core Normal events similar to SN~1994D and SN~2011fe where uniquely detached regions of Si and Ca-rich material are either inferred or detected during photospheric phases \citep{Kasen03,Mazzali05a,Branch05,Smith11,Childress13HVF,Maund13,Childress13}. 

In contrast, those other normal-looking SN~Ia, i.e., SN~2002bo, 2002dj, 2004dt, and 2006X, have broader lines throughout the entire spectrum. Here the ``photospheric velocity features'' of Broad Lined SN~Ia share projected Doppler velocities with``high velocity regions'' of Core Normal SN~Ia; a large window of consistency so far exists for well-matched explosion scenarios (c.f.~\citealt{Maedanature,Maund10a,Dessart14models,Graham1411fe}). Thus, single-component metrics and related classification schemes are limited when the intensity minimum is either momentarily broadened to a trapezoidal-like shape or observed to have a distinctly asymmetric, time-evolving structure arising from, for example, detached or non-standard distributions of \ion{Si}{ii}. 


An alternative yet complimentary classification scheme is to designate events by maximum light expansion velocities. Cases where mean projected \ion{Si}{ii}~$\lambda$6355 Doppler velocities are greater or less than $\sim$~11,800~km~s$^{-1}$ are referred to as high or normal velocity SN~Ia, respectively \citep{WangX09Subtype,WangX13}. Tracing the intensity minimum at 6100~\AA\ has been proven useful for  SN~Ia as cosmological distance indicators \citep{Foley11,Mandel14}. However as for connecting physical diversities with spectral sequences, here too the {\it de~facto} assumption of $\Delta\lambda$/$\lambda_{0}$ for a single line is limited in application for cases where expansion velocities are nearest to so-called ``separation velocities'' for the related subtypes; i.e. the detachment velocity of HV \ion{Si}{ii} in normal velocity Core Normal events like SN~1994D~and~2011fe is a bias for any separation velocity between high and normal velocity subtypes.

\begin{table}
 \centering
 \begin{minipage}{140mm}
  \caption{Hybrid Classifications}
  \begin{tabular}{llcccc}
  \hline
Cross-type& \multicolumn{4}{l}{SN Name\footnote{SN~Ia taken from \citet{Foley12c} and \citet{Blondin12}.}}  \\
\hline
High-velocity & 1996bl & 2001r & 2004bk & 2009ig\\
Core Normal & \\
\hline
Normal-velocity & 1981B & 1989B & 1997E & 1998dh \\
Broad Line & 1999cc & 1999ej & 1999h & 2000cf \\
 & 2000cw  & 2001bf & 2001da & 2001ep \\
 & 2002ha & 2002hd & 2003ch & 2003cq \\
 & 2004as & 2006cj & 2006ev & 2007af \\
 & 2007bm & 2007co & 2007kk  & 2007nq \\
\hline
\end{tabular}
\end{minipage}
\end{table}

Moreover for a given empirical metric, the effective resolution of parameterized \texttt{SYN++} velocities, $\delta$$v$, is approximately 500~km~s$^{-1}$; a SN~Ia with a true mean expansion velocity about 11,900~km~s$^{-1}$ is unlikely distinguishable from another SN~Ia with a true mean expansion velocity about 11,700~km~s$^{-1}$. Subsequently, divisions or boundaries between sub-groupings that are sharply defined provide an unreliable measure of physical diversities.\footnote{Despite differing samples, \citet{Blondin12} find a separation value of 12,200~km~s$^{-1}$, which is 400~km~s$^{-1}$ higher than that found by \citet{WangX09Subtype}, i.e., higher than 11,800 km~s$^{-1}$ by typical $\delta$$v$ of $\sim$~500~km~s$^{-1}$.}

Furthermore, and despite similar overlap among members of various spectroscopic classification schemes, the rate in change of a singular point termed the absorption minimum does not necessarily reflect the true underlying photospheric velocity \citep{Branch77}. Whether or not the radial opacity profile of the ejecta is one of a continuous distribution, two or more distinct components, or a non-standard distribution does not necessarily reflect the rate in change of the associated intensity minima near 5750~\AA\ and 6100~\AA, nor the rate in change of respective pseudo-equivalent widths during photospheric phases; relations may exist but they are not tight. 

In terms of the classification criteria of \citet{Branch06} and \citet{WangX09Subtype}, it is therefore not uncommon to find so-called ``high-velocity Core Normal'' SN~Ia alongside ``normal-velocity Broad Line'' SN~Ia (Table~2). However, this kind of overlap does necessarily imply the existence of an additional branch of spectroscopic diversity. Rather, first-order quantifications of SN Ia through a given compound feature are limited in application for mapping observed and therefore physical diversities.

In fact the recent and nearby SN~2014J has also had conflicting empirical qualifications as a HV SN~Ia, however it too is probably a ``low-velocity gradient,'' Core Normal SN~Ia with a slightly larger than normal blue-shifted absorption trough near maximum light \citep{Marion14}. For SN~2014J, it is likely that either a distinctly separate HV component of \ion{Si}{ii} is affecting the position of the 6100~\AA\ intensity minimum, or initially weak PV~\ion{S}{ii}~$\lambda$6305 is contributing with time as expected \citep{Silverman12spectra,vanRossum12,Folatelli13}. Had a spectrum of SN~2014J been obtained earlier, this might be as apparent as it was for SN~1990N and SN~2012fr \citep{Leibundgut91,Jeffery92,Maund13,Childress13}.

\subsection{Luminosity versus Projected Doppler Velocity}

\begin{figure}
\centering
\includegraphics*[scale=0.45, trim = 0 80 690 0 ]{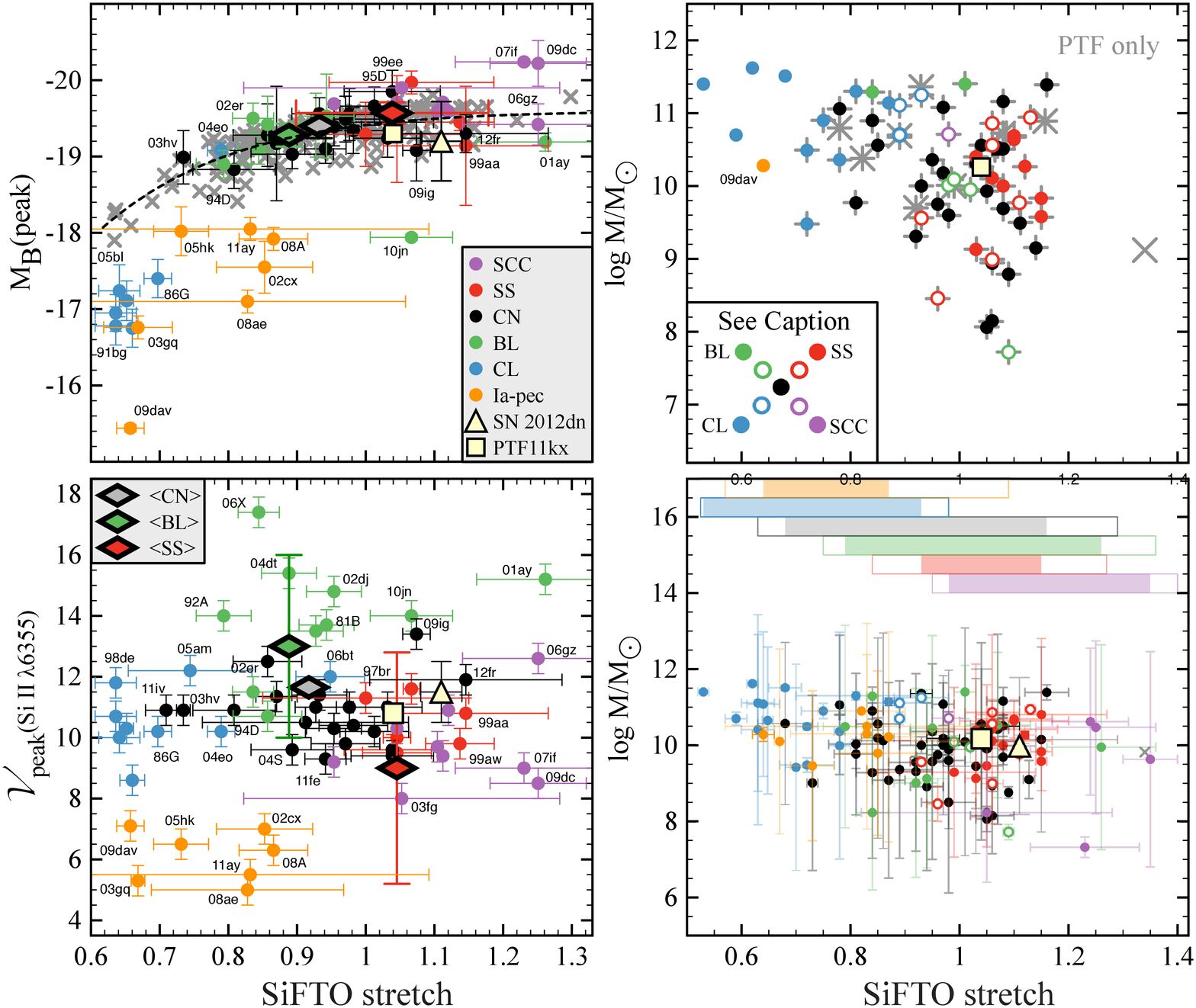}
\caption{[{\it Top}] Peak M$_{B}$ versus \texttt{SiFTO} light curve stretch for literature SN~Ia \citep{Conley08}. See \citet{Parrent14} for references. Grey data points are from \citet{Folatelli12}, \citet{Blondin12}, and \citet{Pakmor13}. Mean values for Core Normal, Broad Line, and Shallow Silicon (diamonds) are shown as reported by \citet{Blondin12}. [{\it Bottom}] Rough estimates of maximum light expansion velocities (assuming a single component of the \ion{Si}{ii}~$\lambda$6355 doublet) versus light curve stretch.}
\label{Fig:diversity}
\end{figure}

The peak luminosity and the shape (or width; hereafter ``stretch'') of SN~Ia light curves have meanwhile served as auxiliary indicators for the state and mass of the ejecta \citep{Stritzinger06,Scalzo14a} in addition to the pre-SN environment and candidate progenitor systems \citep{Sullivan10,Pan14,Pan15}. In general, SN~Ia populate about a width-luminosity relationship. SN~Ia that are more luminous than normal tend to have broader light curves \citep{Khokhlov93,Phillips93,Kasen07,Mandel11}.

In an effort to gauge the intersection between a width-luminosity relationship and the space of nearest-neighbor SN~Ia subtypes \citep{Jeffery07}, in Figure~\ref{Fig:diversity} I have plotted SiFTO light curve stretch values for literature SN~Ia against peak {\it B}-band magnitude and empirically estimated, mean projected Doppler velocities near maximum light. SN~2012dn is shown as a yellow triangle and SN~Ia subtypes are colored according to the pseudo-equivalent width classification of \citet{Branch06}. In the top panel of Figure~\ref{Fig:diversity}, a relationship between the stretch of the light curve and peak brightness is evident, and the overlapping sequence among SN~1984A-likes, 1991T-likes, 1994D-likes, and SN~2012dn-likes is exceptional while exhibiting considerable spread about the width-luminosity relation.

\subsection{On Creating Spectroscopic False-Positives}

Low-order empirical metrics are most susceptible to spectroscopic false-positives. For example, \citet{Pan15} recently implemented the gaussian-line fitting method of \citet{Maguire14} to derive associations between the host environment of each SN~Ia and high velocity features of \ion{Si}{ii} and \ion{Ca}{ii}. In particular, a positive-correlation between singular line-velocities of \ion{Si}{ii}~$\lambda$6355 and logM$_{Stellar}$ was found with $\Delta$$v$$_{Si~II}$~$\sim$~1000~km~s$^{-1}$ between the extremities of binned values for M$_{Stellar}$; as this is only twice the resolution obtainable for SN line-blending via direct analysis (c.f.~Figure~\ref{Fig:converge}), it is therefore worthwhile to examine the method used for spectral line measurements.

In particular, $v_{min}$ for (PV~$+$~HV)~\ion{Si}{ii} and \ion{Ca}{ii} were estimated without accounting for the minimum number of transient lines producing respective compound features; this will have some non-negligible impact on fractional detections and non-detections, but more significantly it will impact estimates of $v_{min, Y}(t)$. Specifically, a four-component empirical metric was used (two for each of the 6100~\AA\ and 8200~\AA\ features). Therefore both lower photospheric velocities and an indication for longer lasting high velocity \ion{Si}{ii}~$\lambda$6355 are partially obscured by an expected increase in line-strength from photospheric \ion{S}{ii}~$\lambda$6305 (c.f. \citealt{HatanoAtlas,Silverman12spectra,Folatelli13}).

Of course these factors may not significantly influence when a given SN~Ia is suspected of having a SN~2012fr-like set of high velocity \ion{Si}{ii} and \ion{Ca}{ii} features. However, without the time-dependent inclusion of at least \ion{S}{ii}~$\lambda$6305, equally time-dependent estimates of \texttt{log~$\tau$} and \texttt{v$_{min}$} for both (PV~$+$~HV) \ion{Si}{ii} solutions will assume whatever data-driven minimization can be obtained. 

How imprecise might PV or HV~\ion{Si}{ii} be if PV~\ion{S}{ii} is omitted? With \texttt{SYN++}, for example, when a fit is made to the ``\ion{S}{ii}~W,'' the $\lambda$6305 line of \ion{S}{ii} effectively straddles the 6100~\AA\ region. If \ion{S}{ii} is omitted from the fit, \texttt{v$_{min}$} for either PV or HV~\ion{Si}{ii} is free to roam between 500~$-$~2000~km~s$^{-1}$ about optimum values depending on the time-dependent strength of \ion{S}{ii} $\lambda$6305 (c.f.~Figure~\ref{Fig:converge}). This is despite the fact that this particular line complex of PV~\ion{S}{ii} is not nearly as dominant as PV~\ion{Si}{ii}~$\lambda$6355 \citep{vanRossum12}. However because these resolutions of projected Doppler velocities are based on data-driven minimization, more well-determined margins of error could be achieved through either detailed modeling or complete observational coverage. 


An additional source of uncertainty in $v_{min}(t)$ can occur when inferring high velocity \ion{Ca}{ii} via the region encompassing the infrared triplet; a four-component metric, i.e., (PV~$+$~HV)~\ion{Si}{ii} and (PV~$+$~HV) \ion{Ca}{ii}, does not account for time-dependent lines of \ion{O}{i}, \ion{Mg}{ii}, and \ion{Fe}{ii} within the loose boundaries of 8200~\AA\ features. In fact for the infrared triplet of \ion{Ca}{ii}, it is not often the case that all three lines are concurrent with conspicuous signatures thereof (\citealt{Thomas04,Silverman13SN2013bh}; see also Fig.~5$-$11 of \citealt{Parrent14}). Instead for each compound feature, one must effectively remove the sum of other contributing species before parameters associated with an ion can be deduced and subsequently measured over time \citep{vanRossum12,Parrent14}. 

Furthermore, in terms of LTE P~Cygni summation, the lack of a conspicuous HV~\ion{Ca}{ii} infrared triplet does not also hold true for HV \ion{Ca}{ii} when considering the relatively stronger Fraunhofer H\&K lines. It may be true that any calcium at high velocities is associated with low abundances thereof. However any lack of relatively strong HV \ion{Ca}{ii} features for SCC SN~Ia, for example, does not yet preclude a plausible association with ``normal SN~Ia.'' This is an important point to consider should either signatures of the infrared triplet of HV~\ion{Ca}{ii} dip below conspicuousness post-discovery or HV~\ion{Ca}{ii} features of SCC SN~Ia mimic fast-evolving 6300~\AA\ signatures found for some other SN~Ia \citep{Zheng13}. Therefore it is suggested that empirical studies of SN spectra increase the number of atomic components from one, two, and four to 11 or more to further improve estimates of project Doppler velocities.
 
\section{Summary}

Pending improved spectrum synthesis calculations from detailed explosion simulations, a wealth of distinct progenitor configurations and explosion scenarios can be found throughout the literature that may not match with the intended SN target after all. It is certainly true that computing exact replicas of time-series spectra from an incinerated white dwarf is a tall order considering the computational resources needed to do so \citep{Hauschildt14}. However, when the spectra of some observed SN are able to better match SN in other galaxies, for example, as ``clones'' \citep{Chakradhari14}, it becomes reasonable to falsify those equivalencies made between a given event and an insufficiently matched model, particularly when a given model may pair better with a separately normal or peculiar event (c.f.~\citealt{Parrent11,Sasdelli14}).

Bearing in mind that spectroscopic interpretations of SN~I are foremost limited by a minimum resolution imposed by line-blending from 11$+$ ions ($\sim$~500~km~s$^{-1}$ for empirical line metrics), generally only statements of {\it ad~hoc} comparative consistencies are accessible per object \citep{Hicken07,Scalzo14,Milisavljevic14}. It is important to emphasize this is the situation whether or not one approaches the problem in its fullest detail, where simplifying assumptions dictate what physical information can be sought after for comparative studies rather than a particular subset of critical ingredients. 

In fact, as current samples have primarily focused on providing boundary conditions at optical wavelengths, a few interpretational discrepancies remain for most subcategories of SN~I (c.f.~\citealt{Parrent07,Dessart12}). Consequently, future datasets comprised of well-observed SN~I may reveal more or fewer connections among those appearing outside the historical breadth of normal and peculiar events.

Fortunately, when lists of active spectral lines and/or atomic species accompany various explosion models, assessing spectral sequences of SN is effectively streamlined \citep{vanRossum12,Dessart12,Blondin13,Friesen14}. If there is an interest in drawing attention to discrepancies between \texttt{SYN++} and more detailed spectrum synthesizers, this can be done within the public forum \citep{Kerzendorf14}. Estimates of projected Doppler velocities and related uncertainties could also benefit from archiving single-ion spectra for all detailed explosion simulations.

Our general understanding of progenitor diversities could also take advantage of well-determined upper-limits on abundances of elements not innately connected to the model or perceived explosion scenario (i.e., H, He, C, N, and O, in general; \citealt{James10,Hachinger12,Jerkstrand15}). A related goal is estimating the radial extent of detectable atomic species, including any considerable overlap of freshly synthesized and unburned progenitor material \citep{Garavini04,Parrent12,Valenti12}.

As studies of supernovae enter a photometrically-based era via, e.g., the Large Synoptic Survey Telescope, it is also important to recognize that some observational strategies looking to enhance our understanding of supernova physics will ultimately lack high-quality spectroscopic follow-up of most discoveries. In particular, should surveys of SN~I be unable to provide complete coverage in wavelength and time, it will be difficult to confidently process and extract few if any of the six so-called observables listed in \S3. 

Therefore, surveys that promptly type and announce newly found supernovae are needed to maximize complete coverage \citep{LCOGT,Sand14}. Upcoming facilities including the Giant Magellan Telescope, the Thirty Meter Telescope, and the European Extremely Large Telescope could also serve well for late-time follow-up of both nearby and higher redshift events. 



\section*{Acknowledgements}

This work was supported by the Las Cumbres Observatory Global Telescope Network. This research used resources of the National Energy Research Scientific Computing Center, which is supported by the Office of Science of the U.S. Department of Energy under Contract No. DE-AC02-05CH11231, and a grant from the National Science Foundation: AST-1211196. 

This work was made possible by contributions to the SuSupect \citep{Richardson01} and WISeREP databases \citep{WISEREP}, as well as David Bishop's Latest Supernovae page \citep{Galyam13}.

Finally, I wish to thank E.~Baron, D.~Branch, M.~Drout, R.~Fesen, B. Friesen, M.~L.~Graham, D.~A.~Howell, B.~Kirshner, D.~Milisavljevic, D.~Sand, A.~Soderberg, S.~Valenti, J.~Vinko, and C.~Wheeler for helpful discussions and comments on previous drafts.  

\bibliographystyle{mn2e}
\bibliography{jparrent_bib}{}

\bsp

\label{lastpage}

\end{document}